\documentclass[final]{svjour2}
\usepackage{graphicx}
\usepackage{rotating}
\usepackage{amssymb}
\usepackage{mathptmx}
\usepackage[numbers]{natbib}
\makeatletter
\journalname{Journal of Low Temperature Physics}

\bibpunct{}{}{,}{s}{}{,}

\begin{document}

\newcommand{\hdblarrow}{H\makebox[0.9ex][l]{$\downdownarrows$}-}
\title{Phonon-Based Position Determination in SuperCDMS iZIP Detectors}

\author{Adam Anderson, for the SuperCDMS Collaboration}

\institute{Department of Physics, Massachusetts Institute of Technology, Cambridge, MA 02139, USA\\
\email{adama@mit.edu}}

\date{07.15.2013}

\maketitle

\begin{abstract}
SuperCDMS is currently operating a 10-kg array of cryogenic germanium detectors in the Soudan underground laboratory to search for weakly interacting massive particles, a leading dark matter candidate. These detectors, known as iZIPs, measure ionization and athermal phonons from particle interactions with sensors on both sides of a Ge crystal. The ionization signal can be used to efficiently tag events at high radius and near the top and bottoms surfaces, where diminished charge collection can cause events to mimic WIMP-induced nuclear recoils. Using calibration data taken with a $^{206}$Pb source underground at Soudan, we demonstrate rejection of surface events of $(4.5 \pm 0.9) \times 10^{-4}$ with 46\% acceptance of nuclear recoils using the phonon signal only. We also show with $^{133}$Ba calibration data underground that the phonon channels can efficiently identify events near the sidewall. This phonon-based approach can also be extended to lower energies than the ionization-based position reconstruction.


\end{abstract}

\section{Introduction}
The SuperCDMS experiment is a dark matter search consisting of 15 ultra-pure cylindrical germanium crystals known as interleaved z-sensitive ionization and phonon detectors (iZIPs) each with a mass 0.62~kg and operated at 50~mK at the Soudan Underground Laboratory in Minnesota. Transition-edge sensors (TESs) and aluminum phonon absorbers called QETs are patterned on the top and bottom surfaces of the crystals and grouped in parallel arrays to form 8 phonon channels. These phonon channels measure non-equilibrium phonons produced by particle interactions in the detector. Electron-hole pairs are also produced by particle interactions, and this ionization energy is measured by electrodes that are interleaved with the phonon sensors. When WIMP scatters off a nucleus, it produces a nuclear recoil as opposed to the electronic recoils produced by the majority of background radiation. At a fixed recoil energy, the fraction of energy deposited in ionization, called the ionization yield, is approximately 0.3 for nuclear recoils (NR), while it is 1 for electronic recoils (ER). The simultaneous measurement of ionization and phonons therefore provides powerful discrimination between well-reconstructed electronic and nuclear recoils.

Background events near the top and bottom surfaces and around the outer sidewall of an iZIP can exhibit ionization yield below 1, increasing the chance of leaking into the NR signal band. Betas at 100~keV and gammas at 10~keV stop at about 50~$\mu$m from the surface, where ionization collection can be lower due to back-diffusion of energetic charges into the wrong polarity electrode \cite{HertelThesis}. Radiogenic nuclei such as $^{206}$Pb from decay of $^{210}$Pb implanted on nearby passive material produce nuclear recoils with ionization yield about 0.2 and a penetration length of microns. Finally, the electric field is weak and fringing near the outer sidewall of the detector, which may cause charge trapping that reduces the yield of all events, including high-energy gammas.

\begin{figure}
\begin{center}
\includegraphics[width=0.75\linewidth,keepaspectratio]{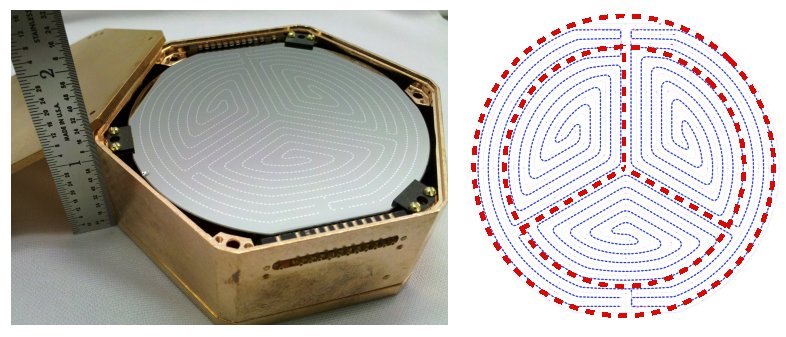}
\end{center}
\caption{(Color online) iZIP detector (left) and the layout of phonon-sensing QETs on each side, with the partitioning into each phonon channel shown in red (right).}
\label{quad}
\end{figure}

The ionization electrode structure of the iZIP efficiently tags potential background events at both high radius and near the surface. An outer guard electrode forms a ring around the outer edge of each detector face, which measures a signal when the event is at high radius. The phonon and charge sensors are biased at 0V and +2V (-2V), respectively, on side 1 (side 2). This arrangement produces an overall drift field that causes electrons and holes to drift to opposite sides of the detector when an event occurs in the bulk. Electrons and holes from events near the surface, on the other hand, both drift to the same side of the crystal. The difference in charge collection between the two sides is therefore a powerful discriminator of surface events \cite{iZIPSE,doughty}.

Since the phonon channels are sensitive to non-equilibrium phonons, some of the position information of the event is contained in the pulse shapes measured by the phonon sensors and the partitioning of energy between the different channels. Phonon-based identification of surface events and sidewall events is advantageous because it provides some redundancy for the charge-based methods. It also has potential to function at lower energies than the charge, since the phonon resolution is better than the charge resolution in iZIP detectors, and charge-based position estimation is worse, by definition, in regions where charge collection is poor. Extending discrimination to lower energies is especially important in searches for light WIMPs, a topic of recent experimental interest \cite{SiPaper}.

\section{Phonon Physics in the iZIP}
When an event occurs in an iZIP, an initial population of high-frequency ($\nu > 1$~THz) prompt phonons are produced. These phonons interact primarily by elastic isotope scattering and inelastic anharmonic decay. The rate for isotope scattering in Ge is given by\cite{LemanMC} $(3.67 \times 10^{-41} \textrm{s}^3)\nu^4$, while the rate for anharmonic decay is given by $(6.43 \times 10^{-55} \textrm{s}^4)\nu^5$. At THz frequencies, anharmonic decay causes rapid downconversion of phonons to lower energies. As they down convert, the isotope scattering and anharmonic decay rates decrease quickly and the mean free path of the phonons increases, resulting in a quasidiffusive propagation of phonons away from the initial interaction location. To see how the position information can be probed, consider that a 0.5~THz phonon has a lifetime of about 40~$\mu$s--much longer than the $1.6\mu$s sampling time of the readout electronics--and the radius of the quasidiffusively expanding phonon cloud is expected to be about 0.2~mm \cite{Wolfe,PyleThesis}  These prompt phonons are preferentially absorbed by sensors near the interactions site since the high rate of scattering can produce many interactions with nearby instrumented surfaces. This results in significant position information in the early time component of the pulse, with nearby sensors measuring higher amplitudes and faster risetimes. As time continues, the phonons continue to downconvert until their mean free path exceeds the size of the crystal in the ballistic regime, at which point all position information is lost.

Position information is also carried by Luke-Neganov phonons \cite{Luke} produced when charges drift across the crystal. When charge carriers in Ge reach their constant drift velocity, additional potential energy released as they drift through the electric field is converted to Luke-Neganov phonons. Since the electric fields are strongest near the top and bottom surfaces of the detector, the Luke phonon power emission is greatest there. Electrons and holes emit Luke phonons at the frequency scales of about 0.3~THz and 0.7~THz, which results in quasidiffusive propagation in a localized cloud near the surface, and strong absorption in a single phonon channel.

\begin{figure}
\begin{center}
\includegraphics[width=0.75\linewidth,keepaspectratio]{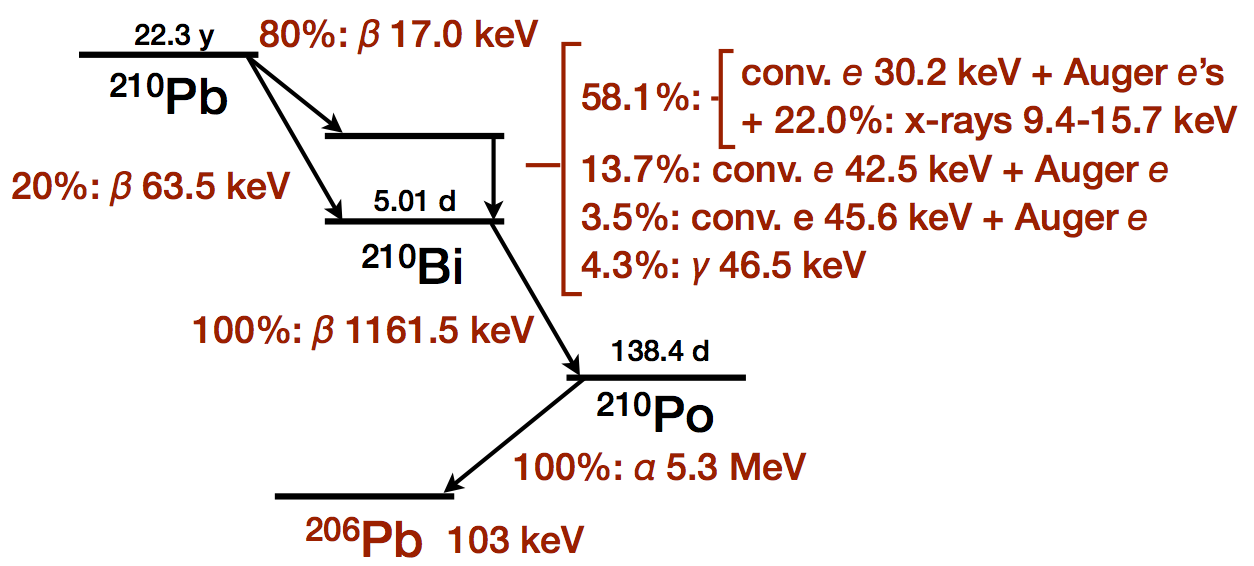}
\end{center}
\caption{(Color online) Relevant decay radiation for the $^{210}$Pb source used in this study.\label{fig:210pbdecay}}
\label{quad}
\end{figure}

\begin{figure}
\begin{center}
\includegraphics[width=0.75\linewidth,keepaspectratio]{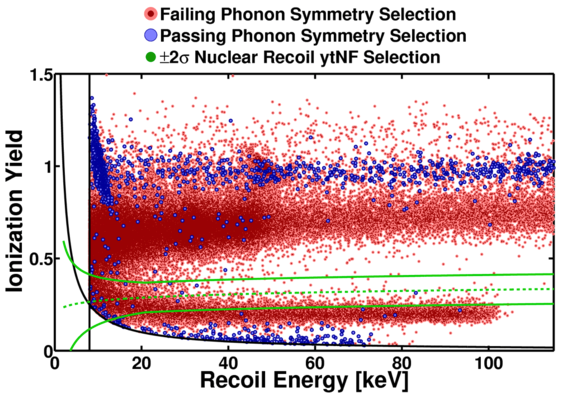}
\end{center}
\caption{(Color online) Low-background data from $^{210}$Pb source in plane of recoil energy and ionization yield.\label{fig:pbdata}}
\label{quad}
\end{figure}

\section{Testing Discrimination}
To test the ability of the phonon signal to identify different types of events, data was taken using two iZIP detectors running in the Soudan Underground Laboratory with $^{210}$Pb sources and a $^{133}$Ba source. The former produce low-energy gammas, betas, and $^{206}$Pb nuclei, as shown in Figure \ref{fig:210pbdecay}, which stop near the surface and are generically classified as ``surface events" that can exhibit reduced ionization yield and contaminate the nuclear recoil signal region. They consist of silicon wafers that were implanted with $^{210}$Pb by exposure to $^{222}$Rn \cite{iZIPSE,edelweiss}. The wafers were positioned above one side of two different detectors in Soudan, and 37.6 live days of unblinded low-background data were analyzed as shown in Figure \ref{fig:pbdata}. The $^{133}$Ba source has strong gamma lines at 302.8~keV, 356~keV, and 383.9~keV, which can produce Compton scatters at high-radius where charge trapping frequently occurs.

For any individual phonon channel or combination of channels, a pulse height estimator is constructed using an optimal filter with a single pulse template. One simple method for quantifying the position-dependence of each event is to compare the relative partitioning of energy between different phonon channels. The resulting phonon energy partitioning will be correlated with the spatial position of the events, so that events near the surface and outer sidewall of the detector can be identified in phonons even if their charge propagation and reconstruction is poor.

\begin{figure}
\begin{center}
\includegraphics[width=0.85\linewidth,keepaspectratio]{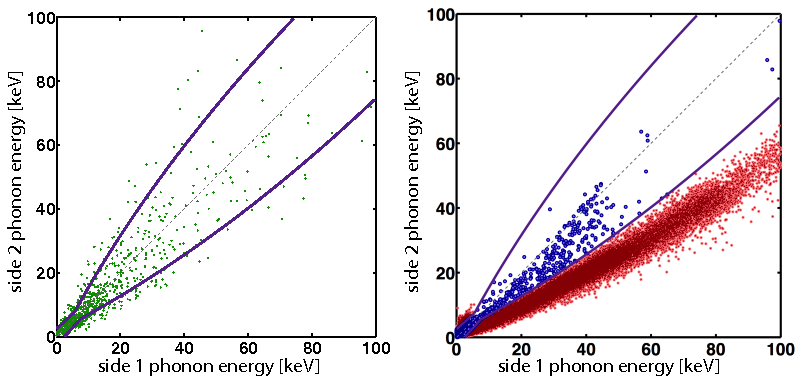}
\end{center}
\caption{(Color online) Phonon energy on side 1 and side 2 of the detector for $^{252}$Cf nuclear recoils (left) and $^{210}$Pb calibration data with ionization yield $< 0.7$ (right). Purple lines indicate an energy-dependent cut optimized to separate surface events from bulk nuclear recoils. Blue (red) points on the right panel are calibration events that pass (fail) the phonon symmetry cut. \label{fig:ps1Vps2}}
\label{quad}
\end{figure}

To tag events near the surface from the $^{210}$Pb source, we can compare the phonon energy measured on side 1 against side 2 of the detector. Traces from all channels on each side are summed and an optimal filter is used to compute the pulse amplitude. The right panel of Figure \ref{fig:ps1Vps2} shows low-yield data from a detector with a source above side 1. These events are overwhelmingly surface events, and they primarily lie in a band below the line of equal energy partitioning. In contrast, nuclear recoils are centered in a broad band with equal energy collected on each side. A cut is constructed in bins of side 1 phonon energy to optimize the signal significance of nuclear recoils at each energy. The resulting cut values are then fit to the functional form shown by the purple lines in Figure \ref{fig:ps1Vps2}; events inside the purple bands are accepted, while events outside are rejected. A small spur of events is also visible above the upper cut boundary at low energy. These events are due to a period of abnormally high electronics noise on one channel, rather than true surface events, and they are ignored in the cut optimization.

\begin{figure}
\begin{center}
\includegraphics[width=0.75\linewidth,keepaspectratio]{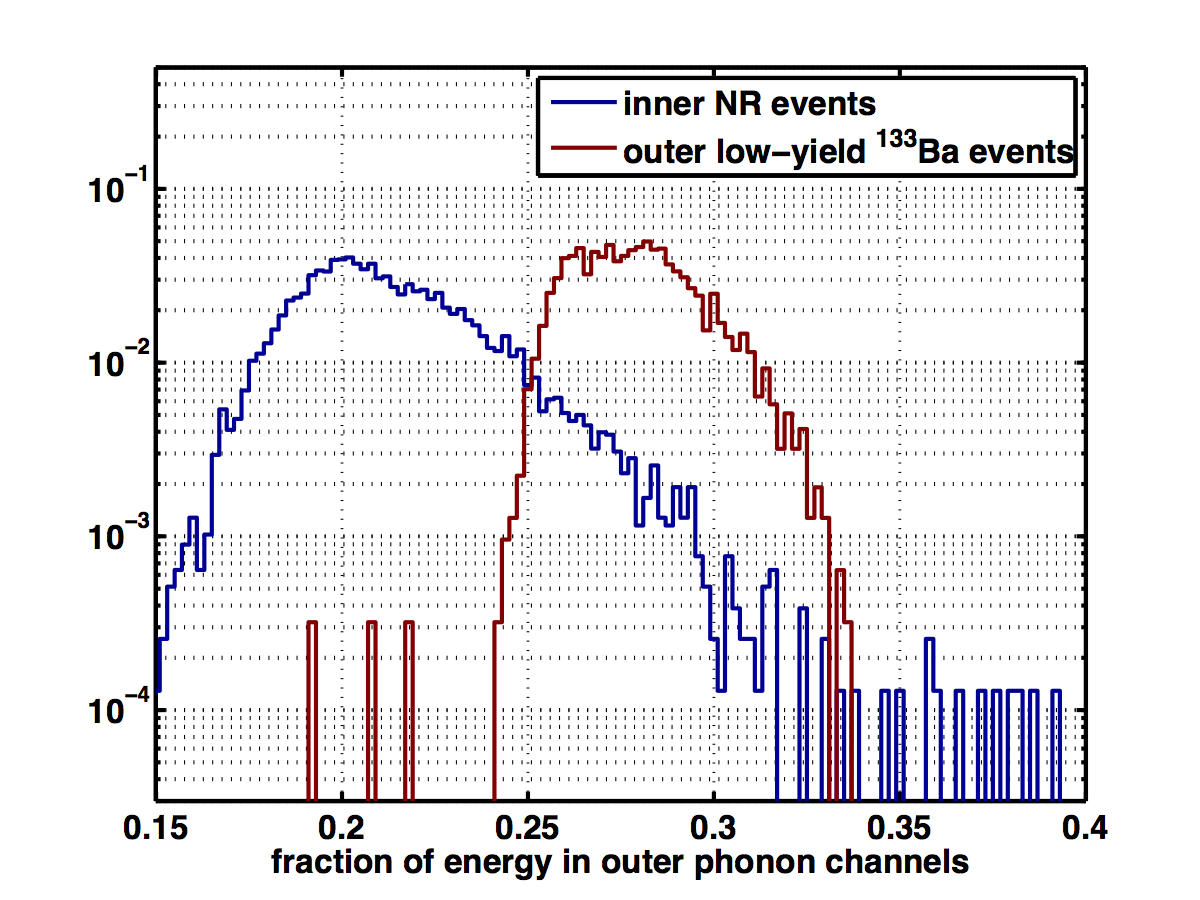}
\end{center}
\caption{(Color online) Fractional energy deposited in the outer phonon channels for nuclear recoils in $^{252}$Cf with charge signal consistent with being in the inner part of the detector (blue), compared with low-yield events from $^{133}$Ba with charge signal consistent with being in the sidewall (red). \label{fig:prpart}}
\label{quad}
\end{figure}

To tag events at high radius in the $^{133}$Ba calibration data, we sum the pulses from the two outer channels and then compare to the total energy of the event to obtain the fraction of energy in the outer channel. Since the calibration gammas consist of a mixture of events near the sidewall and events in the interior of the detector, one must use the charge information to tag high radius events. Figure \ref{fig:prpart} shows the fraction of energy measured in the outer phonon channel for non-surface nuclear recoils in $^{252}$Cf calibration data, compared with low-yield ($<0.7$) electronic recoils in $^{133}$Ba that are tagged as being at high radius by the outer charge channel.

\section{Discrimination Performance}
The optimized cut shown in Figure \ref{fig:ps1Vps2} is able to achieve a surface event leakage of $(4.5 \pm 0.9) \times 10^{-4}$ over an energy range from 8.0 to 115 keV recoil energy with a nuclear recoil passage fraction of 46\% when measured on $^{252}$Cf nuclear recoils. While this is somewhat worse than the surface event rejection achieved using the charge information\cite{doughty}, the phonon surface event discrimination is an independent tag which can complement charge-based methods. It is also extensible to lower energies than the charge-based discrimination, which is a topic of active study within SuperCDMS.

The radial cut is additionally able to tag high-radius charge-suppressed gammas similarly to the charge signal. Cutting events with greater than 24\% of their energy in the outer channels preserves 85\% of the inner nuclear recoils, while rejecting 99.9\% of the low-yield outer events that could potentially leak into the signal region. This shows that even for low-yield events with reduced charge collection, the outer phonon channels can still efficiently distinguish gammas near the sidewall from inner nuclear recoils.

\section{Conclusion and Perspectives}
Phonon propagation in SuperCDMS iZIP detectors operating at Soudan provides strong position sensitivity in regions of the detector which are prone to reduced charge collection that produces leakage into the WIMP-induced NR band. Although the iZIP has excellent charge-based defenses against surface and sidewall events, the simple partition estimators of the phonon energy provides even more discrimination against these events. Phonon-only methods are also extensible to lower energies than charge-based methods and will be used in future low-energy threshold analyses of SuperCDMS data. Time-domain pulse-fitting algorithms for complex pulse shape discrimination\cite{hertelLTD} are also under active development, which may further improve performance over partition-based position estimators.

\begin{acknowledgements}
We thank Julien Billard, Blas Cabrera, Enectal\'i Figueroa-Feliciano, Lauren Hsu, Matt Pyle, and Richard Schnee for valuable discussions. A.A. is supported in part by the Department of Energy Office of Science Graduate Fellowship Program (DOE SCGF), made possible in part by the American Recovery and Reinvestment Act of 2009, administered by ORISE-ORAU under contract no. DEAC05-06OR23100. The SuperCDMS collaboration acknowledges technical assistance from the Soudan Underground Laboratory and Minnesota Department of Natural Resources. The iZIP detectors were fabricated in the Stanford Nanofabrication Facility which is a  member of the National Nanofabrication Infrastructure Network sponsored by NSF under Grant ECS-0335765. SuperCDMS is supported by the US DOE, NSF, NSERC Canada, and MULTIDARK.
\end{acknowledgements}

\pagebreak

\end{document}